\documentclass{article}
\usepackage[utf8]{inputenc}
\usepackage{authblk}
\usepackage{setspace}
\usepackage[margin=1.25in]{geometry}
\usepackage{graphicx}
\graphicspath{ {./figures/} }
\usepackage{subcaption}
\usepackage{amsmath}

\title{Phased Gradient Ultra Open Metamaterials for Broadband Acoustic Silencing}

\author[1,2]{Zhiwei Yang}
\author[1,2]{Ao Chen}
\author[1,2]{Xiaohang Xie}
\author[2,3]{Stephan W. Anderson}
\author[1,2,*]{Xin Zhang}

\affil[1]{Department of Mechanical Engineering, Boston University, Boston, Massachusetts 02215, USA}
\affil[2]{Photonics Center, Boston University, Boston, Massachusetts 02215, USA}
\affil[3]{Department of Radiology, Boston University Chobanian $\&$ Avedisian School of Medicine, Boston, Massachusetts 02118, USA}
\affil[*]{Address correspondence to: xinz@bu.edu}

\date{}

\begin{document}

\maketitle

%%%%%% Abstract %%%%%%
\begin{abstract}
Noise pollution is a persistent environmental concern with severe implications for human health and resources. Acoustic metamaterials offer the potential for ultrathin silencing devices; however, existing designs often lack practical openness and are thereby limited by their functional bandwidths. This paper introduces a novel approach utilizing a phase gradient ultra-open metamaterial (PGUOM) to address these challenges. The PGUOM, characterized by a phase gradient across three unit cells, efficiently transforms incident waves into spoof surface waves, effectively blocking sound while allowing for a high degree of ventilation. Our design provides adjustable openness, accommodates various boundary conditions, and ensures sustained broadband sound insulation. Theoretical, numerical, and experimental validations demonstrate the efficacy of our concept. This innovative approach represents a significant advancement in ventilated acoustic metamaterials, providing both ventilation and high-performance, broadband sound insulation simultaneously. 
\end{abstract}

%%%%%% Main Text %%%%%%

\section{Introduction}
Noise pollution remains a pressing environmental concern with well-documented adverse effects on human health, incurring substantial costs for recovery and mitigation \cite{slabbekoorn2019noise, thompson2022noise}. Various noise control strategies have emerged to address this issue, targeting noise sources, propagation, and receivers, with sound insulation and sound absorption representing fundamental approaches \cite{tao2021recent, gao2022acoustic}. However, traditional designs often rely on materials much thicker than the wavelength of sound waves to achieve the desired phase shift and wave amplitude for effective attenuation \cite{ma2016acoustic}. Additionally, in real-world applications, maintaining ventilation is essential to the core functionality of systems that require noise management, such as air conditioning systems, fans, or drones \cite{liao2021acoustic}. These limitations call for innovative approaches in the realm of ultrathin ventilated silencer.

The advent of metamaterials in the electromagnetic field has led to the development of acoustic metamaterials (AMMs), enabling the creation of ultrathin silencing devices with unprecedented sound control capabilities \cite{liu2014broadband, cummer2016controlling, assouar2018acoustic}. AMMs achieve sound silencing effects through locally resonant unit cells, including membranes \cite{mei2012dark, wang2016membrane}, Fabry-Pérot (FP) resonators \cite{liu2020ultra, zhu2021nonlocal}, Helmholtz resonators \cite{seo2005silencer, zhu2019multifunctional}, and space coiling structures \cite{liang2012extreme, zhang2016three}. To accommodate ventilation, acoustic metamaterials incorporating barriers \cite{wu2018high, xu2020topology, su2022high}, windows \cite{ge2018broadband, ge2019switchable, shi2021ventilative}, metacages \cite{shen2018acoustic, liu2021three}, and sparse arrays \cite{cheng2015ultra, lee2019ultrasparse, gao2022broadband} have been extensively investigated. However, many existing designs struggle to effectively integrate airflow into the sound attenuation process, resulting in a trade-off between openness and attenuation efficiency. Recently, coiled-up structures employing Fano-like destructive interference have shown promise in achieving both excellent ventilation and high sound attenuation \cite{zhang2017omnidirectional, ghaffarivardavagh2019ultra}. However, these designs often impose stringent conditions for sound attenuation across different frequencies, resulting in a narrowed working band.

Phase gradient metasurfaces have garnered significant attention for their unparalleled wavefront manipulation capabilities, enabling various applications such as anomalous refraction/reflection, focusing, beam steering and holography \cite{zhang2014generation, marzo2015holographic, li2018systematic, zhu2018fine, jin2019flat, quan2019passive}. Unlike traditional approaches where frequency selection is critical, phased arrays offer flexibility through the determination of phase distribution and controllable phase shift/amplitude of the responding wave field. Integrating the principles of phased arrays with coiled-up structures holds promise for achieving a broadband ventilated sound silencer.

This paper introduces a phase gradient ultra-open metamaterial (PGUOM), a high-efficiency, broadband, ventilated sound insulator, as shown in Fig. 1. Each super unit cell in the design, composed of three unit cells, forms a phase gradient spanning 2$\pi$, facilitating the transformation of incident plane waves into spoof surface waves, effectively blocking sound while enabling a high degree of ventilation. Here, we analyze and demonstrate how phased arrays achieve broadband ventilated sound silencing, implemented as either a rectangular PGUOM (single super unit cell or multiple super unit cells), or as a cylindrical PGUOM. By selecting the appropriate phase gradient, the openness of the design becomes adjustable. Furthermore, with the aid of length-varying straight barriers and waveguides, we validated our design numerically and experimentally. The proposed design offers a novel approach to ventilated acoustic metamaterials, providing a unique solution that simultaneously achieves effective ventilation and sound insulation. 

\begin{figure}[h]
\centering
\includegraphics[width=\linewidth]{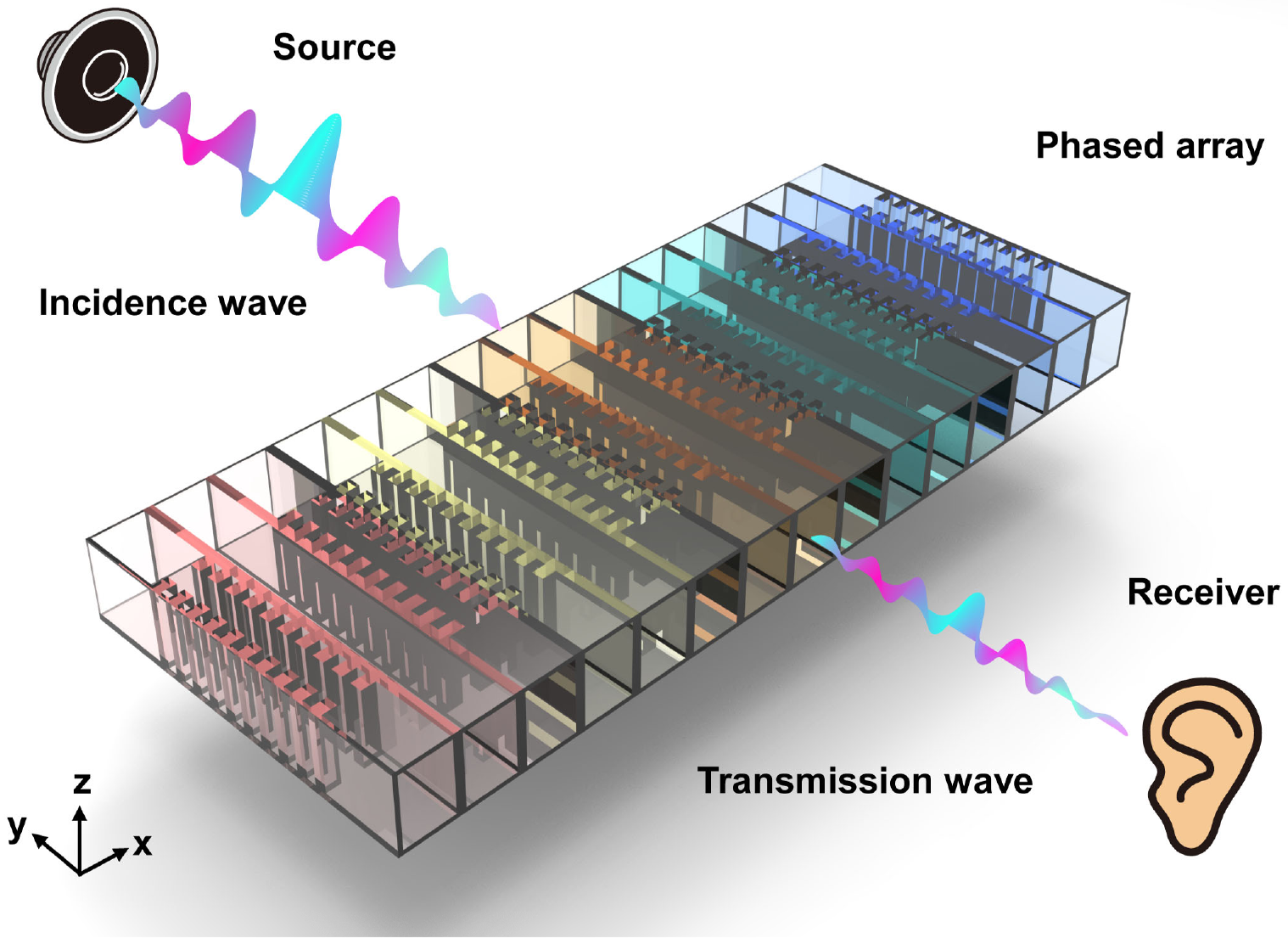}
\caption{Illustration of the phase gradient ultra-open metamaterial (PGUOM) demonstrating simultaneous ventilation and broadband sound ventilation. The PGUOM comprises of five super unit cells, each containing three unit cells with a phase gradient.}
\label{fig:1}
\end{figure}

\section{Phase gradient ultra-open metamaterial (PGUOM)}
We begin by demonstrating the effectiveness of the phase gradient design in achieving an ultra-open metamaterial silencer. Considering an incident acoustic plane wave perpendicularly interacting with a super unit cell comprising three individual unit cells, as shown in Fig. 2(a), with dimensions of height $H$, width $W$, and thickness $t$. Each unit cell, denoted by width $W_i$, possesses distinct acoustic properties characterized by the transmission coefficient amplitude $A_i$ and phase shift $\varphi_i$. Acoustic rigid boundaries are employed to separate adjacent unit cells, thus preventing cross-coupling.

\begin{figure}[h]
\centering
\includegraphics[width=\linewidth]{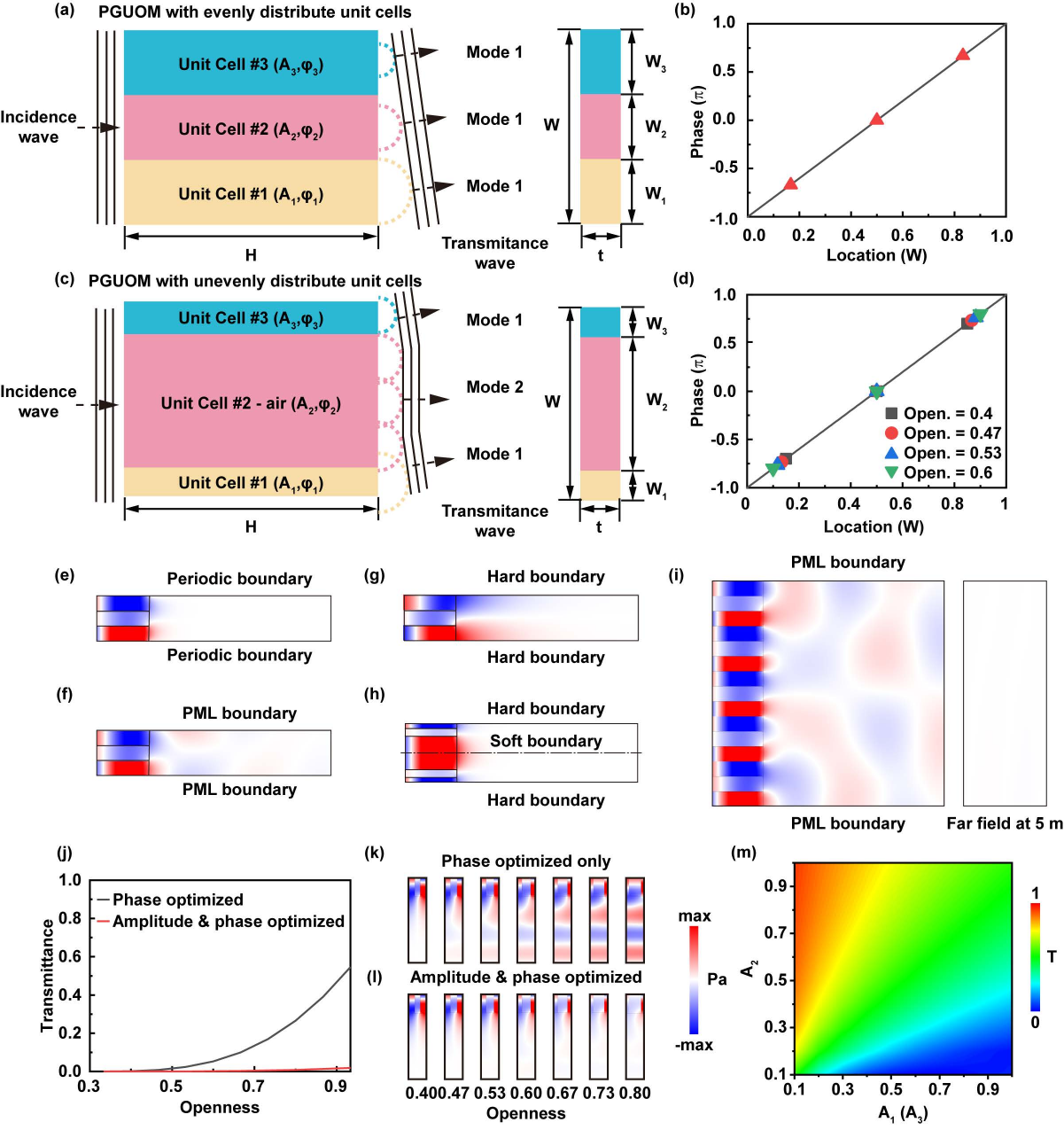}
\caption{Analysis of the PGUOM. (a) Illustration of the PGUOM with evenly distributed unit cells. (b) Phase gradient (black line) and distribution of the three unit cells (red dots) within a super unit cell. (c) Illustration of the PGUOM with unevenly distributed unit cells. (d) Phase gradient (black line) and distribution of the three unit cells (markers) within a super unit cell with different openness levels. (e-i) Pressure field distribution of rectangular PGUOM and cylindrical PGUOM at target frequencies under periodic boundary, PML boundary, hard boundary and soft boundary conditions. (j) Transmittance of the PGUOM under different openness levels at 2000 Hz with phase optimized only (black line) and with both amplitude and phase optimized (red line). (k, l) Pressure filed distribution of the PGUOM under different level of openness at 2000 Hz with phase optimized only and with both amplitude and phase optimized, respectively. (m) The transmittance as a function of amplitude of unit cell 1/3 ($A_1$/$A_3$) and unit cell 2 ($A_2$) at 2000 Hz when the openness is 0.87.}
\label{fig:2}
\end{figure}

When the amplitudes of each unit cell are close and their phases form a phase gradient, the behavior of the transmission wave follows the generalized Snell’s Law:

\begin{equation}
\sin(\theta_{tr})=\frac{d\varphi(x)}{dx}\frac{\lambda_{tr}}{2\pi}=\frac{\lambda_{tr}}{W}\frac{\varphi}{2\pi}
\end{equation}

where $\theta_{tr}$ represents the transmission angles, $\lambda_{tr}$ denotes the wavelength of the transmission waves, $d\varphi(x)/dx$ describes the phase distribution function and $\varphi$ represents the cumulative phase shift of the transmission wave within the super unit cells. If $\sin(\theta_{tr})$ exceeds 1, the transmission wave transforms into a spoof surface wave, bounded at the interface, resulting in minimal amplitude. To achieve this transition, two conditions must be met: first, the phase along a single super unit cell should span 2$\pi$, and second, the total length of the super unit cell should be less than $\lambda_{tr}$. 

Fig. 2(b) illustrates a super unit cell consisting of three evenly distributed unit cells, each meeting the required conditions. The black line illustrates the phase gradient used, $(d\varphi(x)/dx=2\pi/75 (rad/mm))$, under a reference frequency of 2000 HZ. The red dots on the figure represent the specific transmission wave phase and locations of each unit cell. According to equation (1), the value of $\sin(\theta_{tr})$ is 2.29 and significantly exceeds 1, indicating that the transmission angle approaches 90\textdegree{} and the transmission wave becomes a spoof surface wave. When the unit cells within a super unit cell are unevenly distributed (Fig. 2(c)), as long as the phase gradient is preserved (Fig. 2(d)), the incidence wave can still be partially modulated. To validate our theoretical analysis, we employed COMSOL Multiphysics software to model the effect of phased arrays on the transmission wave. We designed ports with specific amplitude and phase, obtained the transmission spectrum and pressure field distribution to assess the transmission response of the design. Perfectly matched layer (PML), periodic boundary, and hard boundary are applied respectively to represent various boundary conditions, details about the simulations can be found in supporting information.

In Fig. 2(e), we present the simulated pressure field distribution of the transmission wave of a rectangular PGUOM at 2000 Hz with periodic boundary conditions  representing ideal phase gradient. The pressure of the incident wave experiences a substantial decrease after passing the design compared to its peak value within the design, indicating a sharp decay of the transmission wave in its initial propagation direction, with most of its energy confined within the design. Additionally, at the interface of the design, the pressure alternates between maximum and minimum values along the lateral direction, further confirming the transmission wave becomes a spoof surface wave, consistent with our analytical results. It's worth noting that, according to equation 1, when the width of a super unit cell is small, there are many combinations of frequency (wavelength) and design parameters for the PGUOM that can still result in a value of $\sin(\theta_{tr})$ larger than 1. Therefore, the design’s working band can be broadened if the conditions are also satisfied at other frequencies. Furthermore, this PGUOM design incorporates a blank unit cell filled with air, positioned at the center of the super unit cell with a certain phase shift, demonstrating its ventilation capabilities when the phased array is used as a broadband sound silencer.

In addition to periodic boundary conditions, we performed simulations with PML boundary conditions, hard boundary conditions on both sides and a combination of hard boundary on one side and soft boundary on the other side (realized through a cylindrical PGUOM, as will be further illustrated). Fig. 2(f-h) show the simulated pressure field distribution for PGUOMs under these boundary conditions. Similar phenomena are observed, where the energy of the transmission wave enhances in the near field and diminishes in the far field, indicating that the transmission wave transforms into a spoof surface wave. Compared to periodic boundary conditions (representing the ideal case of an infinite phase gradient), hard boundaries and soft boundaries slightly reduce the silencing effect, though the effect remains present. This can be attributed to the unique properties of the spoof surface wave. Both soft boundary and hard boundary can induce a superposition of two identity waves. When two spoof surface waves interact horizontally, since they only propagate along the surface, their resultant wave remains confined to the surface without significant vertical leakage. Fig. 2 (i) illustrates the simulated pressure field distribution of rectangular PGUOM composed of five super unit cells. The pressure decreases as the wave passes through the design, and eventually approaches zero in the far field, at a distance of approximately 5 meters. The flexibility of the boundary conditions enriches the applications of the PGUOM in sound silencing. To note, for cylindrical PGUOM, the cross-sectional areas of each unit cell are used instead of the radius when considering the phase distribution function.

PGUOM also offers the advantage of adjustable openness while remaining effective silencing effect. We characterized the openness of the design as $openness=W_2/W$, under the fixed condition $W_1=W_3$. In Fig. 2(c), we present an example of a rectangular PGUOM with unevenly distributed unit cells. To sustain the required phase gradient, as depicted in Fig. 2(d), the phases of the transmission wave and the length of the three units must satisfy the following relationships:

\begin{equation}
\varphi_{1}=-\frac{Openness+1}{2}\times\pi+\varphi_2
\end{equation}
\begin{equation}
\varphi_{3}=\frac{Openness+1}{2}\times\pi+\varphi_2
\end{equation}

According to equation (3-4), the phase of unit cell 1 or 3 is related to the openness of the design and the phase of unit cell 2. Increasing the openness requires an increased phase difference between unit cell 1 or 3 and unit cell 2. 

When the phase of each unit cell within a super unit cell forms a phase gradient and the amplitudes are all equal to 1, the simulated transmittance of the super unit cell and corresponding pressure field distributions under different levels of openness can be obtained. The black line in Fig. 2(j) represents the simulated transmittance ($S_{21}^2$) of the design at target frequency (2000 Hz) when the openness increases from 0.33 to 0.93, and Fig. 2(k) depicts the pressure field distributions when the openness is 0.4, 0.47, 0.53, 0.60, 0.67, 0.73 and 0.80 at the target frequency. When the openness is 0.33, the transmittance approaches 0, and the transmission wave exhibits clear characteristics of the spoof surface wave  (Fig. 2(e)). However, as the openness gradually increases, the transmission wave begins to resemble a plane wave rather than a spoof surface wave (Fig. 2(k)), causing the transmittance to rise to 0.55 and reducing the silencing effect.

However, if the amplitude of each unit cell is also optimized, the silencing effect can be substantially improved. Table 1 provides the optimal amplitudes of unit cells under different levels of openness, determined by targeting the lowest possible transmittance. The red line in Fig. 2(j) represents the simulated transmittance of the design with optimized amplitudes, and Fig. 2(l) illustrates the corresponding pressure field distributions. Although the transmittance still increases with openness, it remains much lower comparing to the case where the amplitudes are all one (black line). The transmittance remains at 0.05 even when the openness reaches 0.93. Moreover, the transmission wave remains a spoof surface wave at high openness levels. Fig. 2(m) plots the transmittance of the design with different amplitudes of the unit cells at an openness of 0.87. The minimum transmittance (blue area) occurs when the amplitude of unit cell 1 or 3 ($A_1 or A_3$) is large, while the amplitude of unit cell 2 ($A_2$) is small.

\begin{table} [b]
\caption{Dimensions, transmission wave phases and optimized amplitudes of the unevenly distributed unit cells of PGUOM with different openness}
\centering
\begin{tabular}{cccccccc}
\hline
$W_1/W_3/mm$ & $W_2/mm$ & $\varphi_1/\pi$ & $\varphi_2/\pi$ & $\varphi_3/\pi$ & $A_1/A_3$ & $A_2$ & openness \\
\hline
25 & 25 & 0.335 & & -0.335 & & 1 & 0.33\\
22.5 & 30 & 0.30 & & -0.30 & & 0.9 & 0.40\\
20 & 35 & 0.265 & & -0.265 & & 0.8 & 0.47\\
17.5 & 40 & 0.235 & & -0.235 & & 0.7 & 0.53\\
15 & 45 & 0.2 & 1 & -0.2 & 1 & 0.6 & 0.6\\
12.5 & 50 & 0.165 & & -0.165 & & 0.5 & 0.67\\
10 & 55 & 0.135 & & -0.135 & & 0.4 & 0.73\\
7.5 & 60 & 0.1 & & -0.1 & & 0.3 & 0.8\\
5 & 65 & 0.065 & & -0.065 & & 0.2 & 0.87\\
\hline
\end{tabular}
\label{tab:1}
\end{table}

This can be explained by the Huygens–Fresnel principle. As depicted in Fig. 2(a, c), when the width of each unit cell is the same, the wavefront of the transmission wave is simple and follows the generalized Snell’s law. However, when the widths of unit cell are different, the wavefront becomes a superposition of two plane waves. One wave (mode 1) follows the generalized Snell’s law while the other one (mode 2) only depends on the phase shift of unit cell 2. Both the amplitude and width of each unit cell affect the intensity of these two modes. To suppress the intensity of mode 2 and improve the performance, the amplitude of unit cell 2 should be small when its width is large.

\begin{figure}[h]
\centering
\includegraphics[width=.9\linewidth]{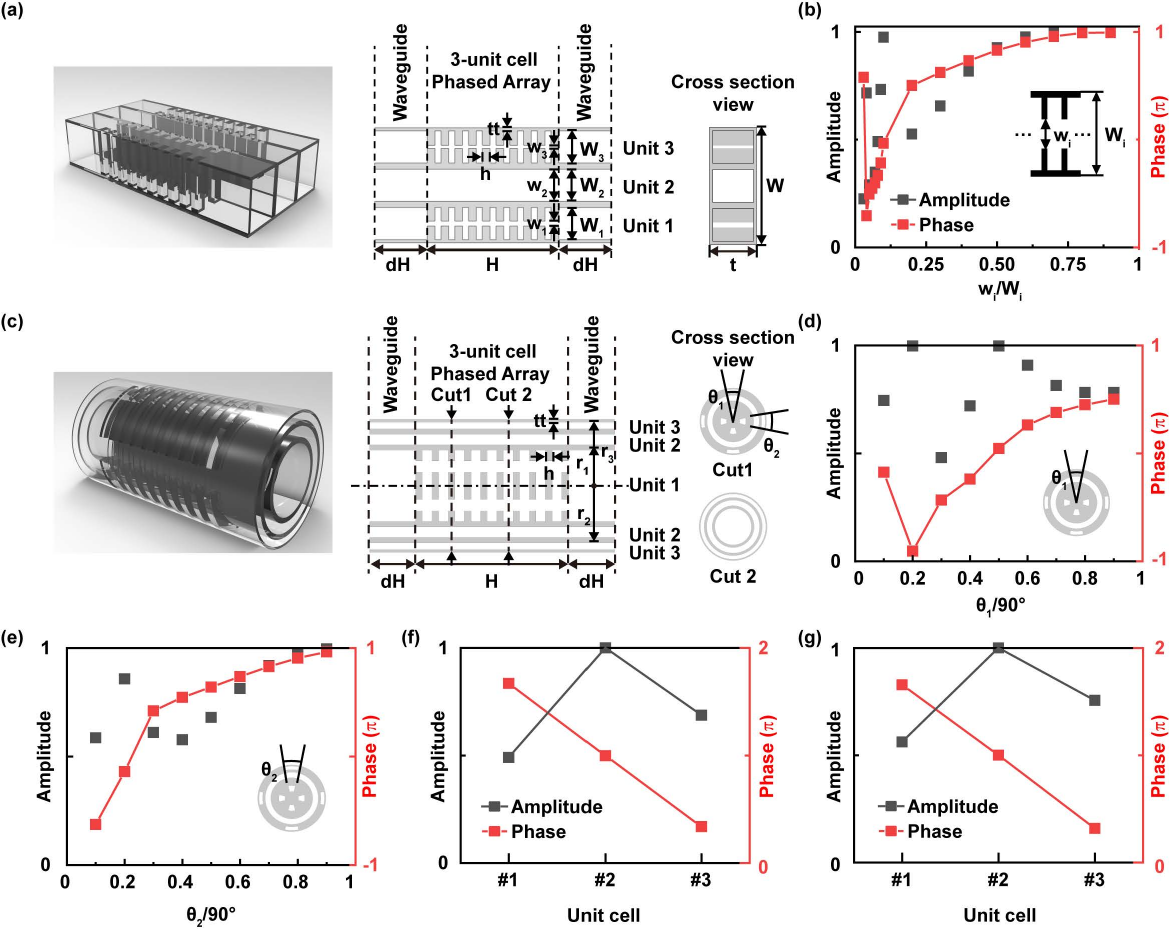}
\caption{(a) Schematic diagram of the rectangular PGUOM design based on the length-varying straight barriers. Parameter values: width of slit in each unit cell $w_1$= 2 mm, $w_2$= 25 mm, and $w_3$= 3.25 mm, height of the silt $h$= 5 mm, number of the barriers n = 10, thickness of the wall $tt$= 2 mm, width of each unit cell $W_1$=$W_2$=$W_3$= 25 mm, width of the phased array $W$= 75 mm, height of the phased array $H$= 85.75 mm, thickness of the phased array $t$= 30 mm, height of the waveguide $dH$= 34.3 mm. (b) Amplitude (black dots) and phase (red solid line) of the transmission wave at 2000 Hz with different $w_i/W_i$. Inset: illustration of a single element of the barriers. (c) Schematic diagram of the cylindrical PGUOM design based on the length-varying straight barriers. Parameter values: angle of the silt within unit cell 1 and unit 3: $\theta_1$= 31.5\textdegree{}, and $\theta_2$= 23.4\textdegree{}, height of the silt $h$= 5 mm, number of the barriers n = 10, radii of unit cell 1, unit cell 2 and unit cell 3: $r_1$= 28.72 mm, $r_2$= 40.62 mm, $r_3$= 49.75 mm, thickness of the wall $tt$= 2 mm, height of the phased array $H$= 114.33 mm, height of the waveguide $dH$= 34.3 mm. (d, e) Amplitude (black dots) and phase (red solid line) of the transmission wave at 1500 Hz with different slit angle in unit cell 1 and 3 respectively. Inset: Cross-section view of the phased array region. (f, g) Amplitude (black solid line) and phase (red solid line) of the transmission wave for different unit cells of the rectangular PGUOM (2000 Hz) and cylindrical PGUOM (1500 Hz) respectively.}
\label{fig:3}
\end{figure}

\begin{figure} [h]
\centering
\includegraphics[width=\linewidth]{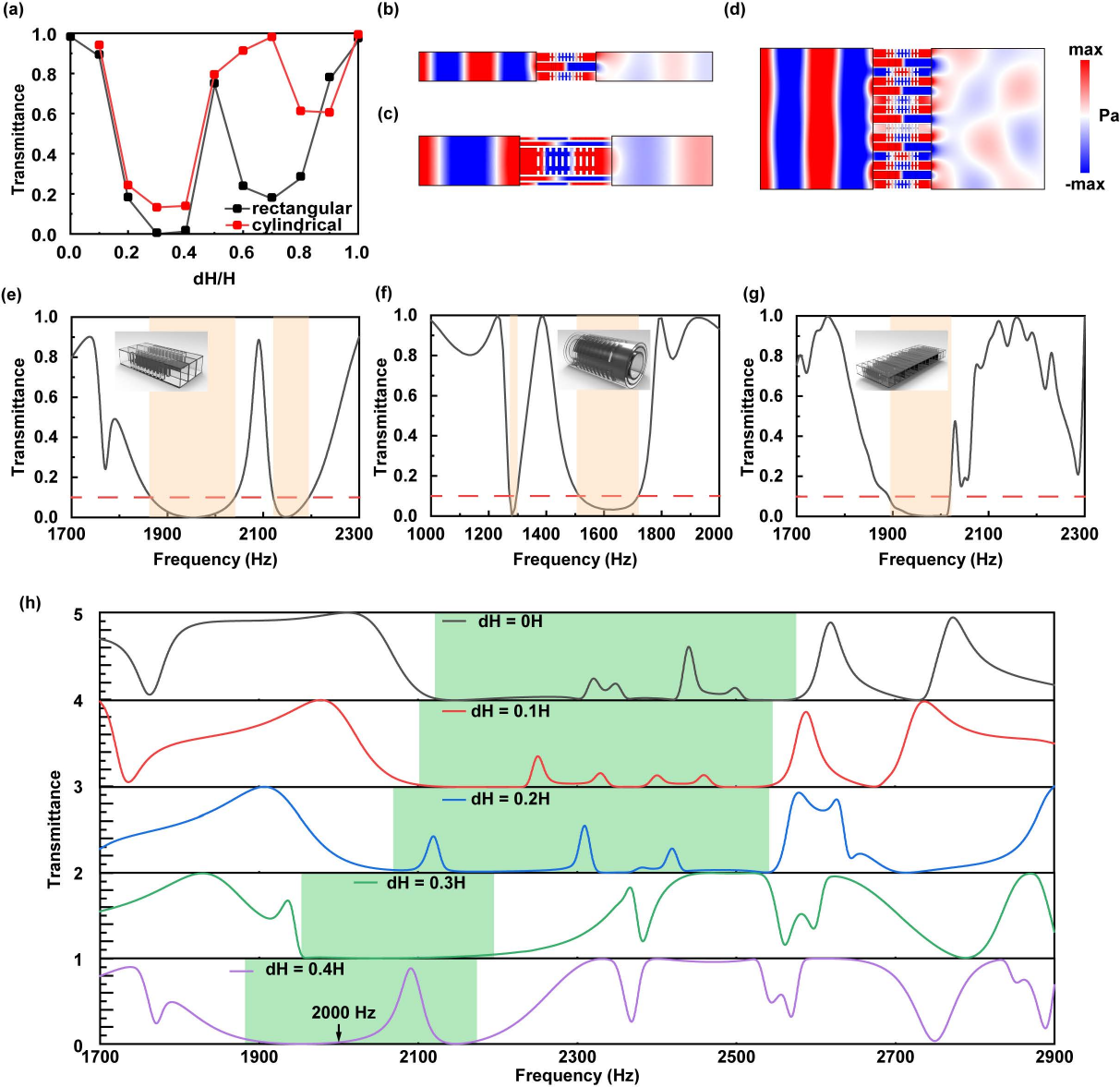}
\caption{(a) Transmittance of the rectangular PGUOM (black dash line, 2000 Hz) and cylindrical PGUOM (red dash line, 1500 Hz) with waveguide’s height ranging from $0.1H$ to $H$. (b-d) Pressure field distribution of the rectangular PGUOM (2000 Hz), and cylindrical PGUOM (1500 Hz) with extra waveguides. (e, g) Transmittance spectra of the rectangular PGUOM across a frequency range from 1700 Hz to 2300  consisting of single super unit cell or five super unit cells. (f) Transmittance spectra of the cylindrical PGUOM across a frequency range from 1000 Hz to 2000 Hz. (h) Transmittance of the rectangular PGUOM with waveguide height ranging form 0 to $0.4H$ across frequencies from 1700 Hz to 2900 Hz.}
\label{fig:4}
\end{figure}

\section{Design validation}
To validate our PGUOM designs, we utilized length-varying straight barriers to achieve the required amplitude and phase by adjusting the ratio of the slit width to the unit cell width ($w_i/W_i$) \cite{fan2023ultrabroadband}, as shown in the inset of Fig. 3(b). For rectangular PGUOM, as shown in Fig. 3(a), we evaluate its performance under a reference frequency of 2000 HZ. The width of each unit cell is set to $W_i$= 25 mm, ensuring the width of the super unit cell ($W$= 75 mm) is smaller than the wavelength under this condition ($\lambda_{tr}$= 171.5 mm). Since we use a blank unit cell at the position of unit cell 2 for ventilation, the height of the super unit cell is set to half the wavelength ($H$= 85.75 mm), yielding an amplitude and phase of unit cell 2 at $A_2=1$ and $\varphi_2=\pi$. The thickness of the wall, the barriers, and the super unit cell are set to $tt$= 2 mm, $h$= 5 mm, and $t$= 30 mm based on fabrication consideration. Fig. 3(b) depicts the amplitude and phase of the transmission wave of a unit cell with different slit widths at 2000 Hz. As the ratio of the slit width to the unit cell width increases from 0 to 1, the phase spans 2$\pi$ while the amplitude varies. When the ratio approaches 1, the unit cell resembles a blank unit cell, with the amplitude close to 1 and the phase close to $\pi$. The slit widths within unit cell 1 and 3 are set to $w_1$= 2 mm and $w_3$= 3.25 mm, respectively, resulting in the following amplitudes and phases: $A_1$= 0.49, $\varphi_1$= 1.67$\pi$, $A_3$= 0.69, and $\varphi_3$= 0.34$\pi$, as shown in Fig. 3(f). The phases of the unit cells (red line in Fig. 3(f)) form a phase gradient of $0.66\pi/25 (rad/mm)$ and the amplitude of the unit cells (black line in Fig. 3(f)) are acceptable (see supporting information for details), ensuring the generalized Snell's Law remains fulfilled for generating a spoof surface wave.

The cylindrical PGUOM is transformed from the rectangular configuration by rotating the super unit cells along its side in the propagation direction. This arrangement creates a cylindrical structure that preserves the phase gradient characteristics of the rectangular design. To achieve the desired phase gradient, we used modified length-varying straight barriers. Additionally, we opted for a fan-shape slit rather than a circular ring slit, as the fan shape provides a larger size for the same area and imposes fewer demands on 3D printing resolution, as shown in Fig. 3(c). Due to the limitation of our measurement setup, the target frequency is set to 1500 Hz as reference, and the outer radius of the design is fixed at 49.75 mm (details can be found in the supporting information). In order to maintain the condition that the area of each unit cell should be equal, the radii of the three layers are set to $r_1$= 28.72 mm, $r_2$= 40.62 mm, and $r_3$= 49.75 mm. The height of the super unit cell is set to half the wavelength as well, $H$= 114.33 mm, to fix the amplitude and phase of unit cell 2. The thickness of the wall, the barriers, and the super unit cell are set to $tt$= 2 mm, $h$= 5 mm, and $t$= 30 mm based on fabrication consideration. Fig. 3(d, e) illustrates the phase and amplitude of the transmission wave at 1500 Hz for different angles of the fan-shaped slits within unit cells 1 and 3. As the ratio increases, the phases of unit cell 1 and 3 increase, while the amplitudes maintains larger than 0.5. Different from the results in Fig. 3(b), the phase does not span 2$\pi$ as $\theta_1$ increases from 0\textdegree{} to 90\textdegree{}. This limitation is due to the structure of unit 1: as the angle of the fan-shape grows, the maximum possible ratio of the slit area to the total area of unit cell 1 is capped at 0.35. Only if this ratio could reach 1 would the phase reach $\pi$, thus limiting the phase range within the unit cell 1. To accommodate phase gradient requirements, we set $\theta_1$/90\textdegree{}= 0.35 and $\theta_2$/90\textdegree{}= 0.26, resulting in the following transmission amplitudes and phases for the three unit cells: $A_1$= 0.56, $A_2$= 1, $A_3$= 0.76,$\varphi_1$= -0.34$\pi$,$\varphi_2$= $\pi$, and $\varphi_3$= 0.32$\pi$, respectively, as shown in Fig.3(g). The phases form a gradient, and the amplitudes are acceptable.

When the three unit cells are combined, the silencing effect is week and diverges from the theoretical results, as shown in Fig. 4(a). Impedance mismatches within unit cells, along with reflections from adjacent unit cells, lead to deviations in the transmission amplitude and phase, reducing the silencing effect. Additionally, the varying amplitudes among each of the designed unit cell further diminish the performance. To improve the silencing effect of the PGUOM, additional waveguides are introduced on both sides of each unit cell. Fine-tuning the waveguide length allows for optimizing the transmission amplitude of each unit cell \cite{zhu2018fine}, thereby achieving a more effective silencing effect. As shown in Fig. 4(a), the transmittance of the rectangular PGUOM and cylindrical PGUOM at target frequencies (2000 Hz for rectangular PGUOM and 1500 Hz for cylindrical PGUOM) with waveguides of varying heights are simulated. The transmission changes with increasing waveguide height, and the optimal heigths ($dH$) are selected as $0.4H$ for the rectangular PGUOM and $0.3H$ for the cylindrical PGUOM, as these configurations yielded the lowest transmittance. Fig. 4(b-d) illustrates the simulated pressure field distribution of the rectangular PGUOM (with both single and an array of five super unit cells) and the cylindrical PGUOM at their target frequencies. In each case, the pressure of the transmission wave decreases significantly after passing through the design, with a spoof surface wave observed at the interface. The excellent sound silencing effect demonstrated that our designs are useful for small-scale, large scale, or cylindrical applications.

\begin{figure}[h]
\centering
\includegraphics[width=\linewidth]{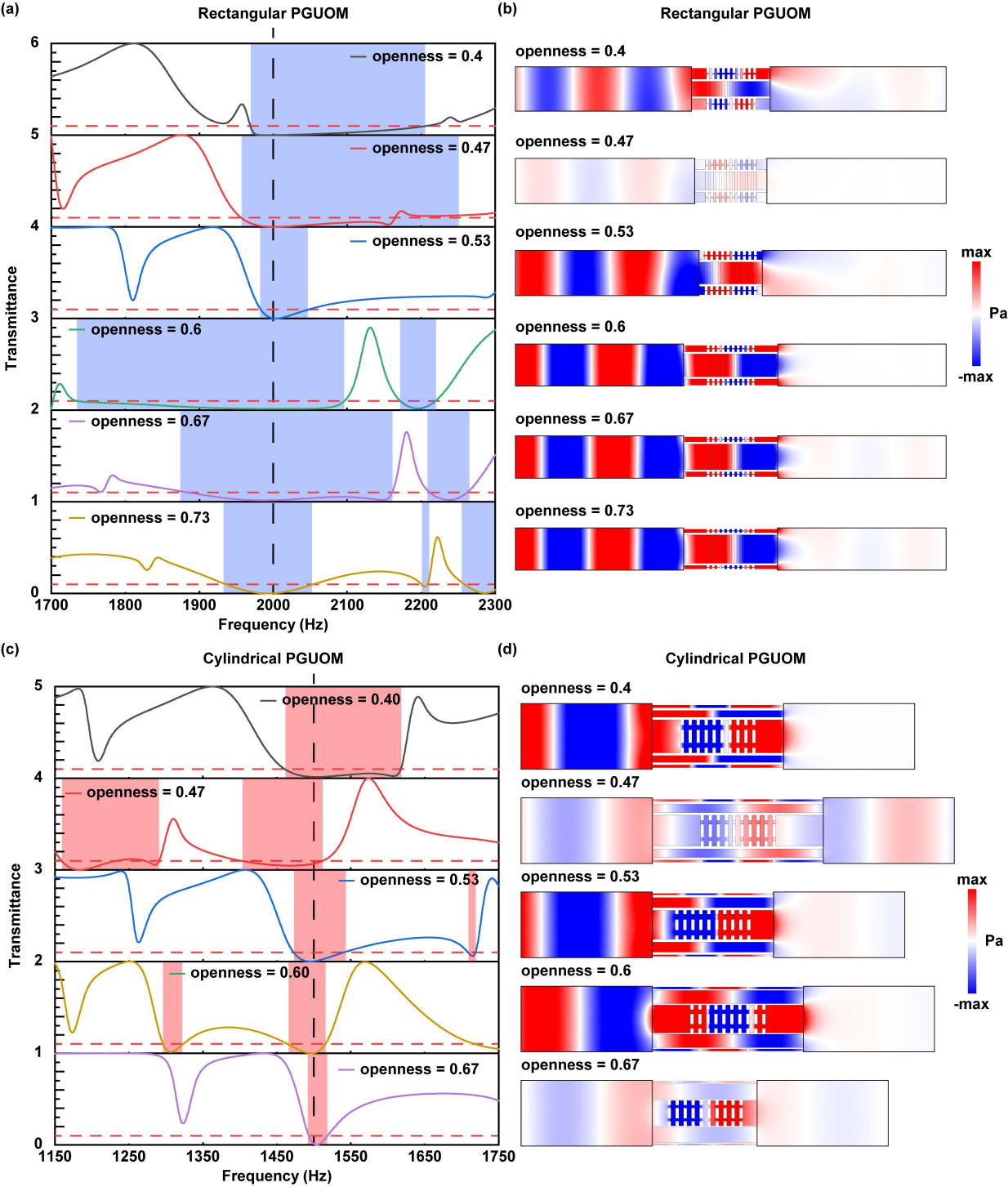}
\caption{(a) Transmittance spectra of the rectangular PGUOM with different levels of openness across a frequency range from 1700 Hz to 2300 Hz. (b) Pressure field distribution of the transmission wave of the rectangular PGUOM at 2000 Hz with different openness levels. Openness of the PGUOM: 0.4, 0.47, 0.53, 0.6, 0.67, and 0.73. Length of the waveguide dH for different openness levels: $0.3H$, $0.2H$, $0.1H$, $0.5H$, $0.5H$, and $0.5H$. (c) Transmittance spectra of the cylindrical PGUOM with different openness levels across a frequency range from 1150 Hz to 1750 Hz. (d) Pressure field distribution of the transmission wave of the cylindrical PGUOM at 1500 Hz with varying levels of openness. Openness of the PGUOM: 0.4, 0.47, 0.53, 0.6, and 0.67. Length of the waveguide dH for different levels of openness: $0.38H$, $0.63H$, $0.3H$, $0.5H$, and $0.19H$.}
\label{fig:5}
\end{figure}

Next, we explore the broadband properties of the designs by simulating the transmission spectra of the rectangular PGUOM and cylindrical PGUOM over frequency ranges of 1700-2300 Hz and 1000-2000 Hz, respectively, as shown in Fig. 4(e-g). A transmittance of 0.1 is set as the threshold for assessing the silencing effect. For rectangular PGUOM with a single super unit cell, the transmittance spectrum show dips at 1950 Hz and 2150 Hz, with effective silencing bands between 1860-2050 Hz and 2120-2200 Hz. For the rectangular PGUOM with five super unit cells, the spectrum shows a dip from from 1890 Hz to 2040 Hz. The cylindrical PGUOM spectrum has a sharper dip between 1270-1300 Hz and a broader dip between 1500-1730 Hz. The broadband property arises from the robust requirement of spoof surface wave generation. Around the target frequencies, the phase gradients are approximately maintained, and the amplitudes are within an acceptable range (see supporting information for details), leading to a decrease in transmittance, though not to the ideal level. Fig. 4(h) shows the transmission spectra of the rectangular PGUOM with varying waveguide heights. As the waveguide height ($dH$) increases, the effective silencing band shifts to lower frequencies, enabling sound silencing at difference target frequencies by adjusting waveguide height ($dH$).

We also demonstrate that our deign achieves sound silencing under varying degrees of openness. Based on the openness level, we first calculated the required phased distribution and dimensions of each unit cell using equation (3) and (4). The slit within each unit cell was then adjusted to achieve the desired phase. Finally, waveguides were added to both sides of each unit cell to enhance silencing at the target frequency. Further optimizations of each unit cell’s dimensions and waveguide lengths were conducted based on simulation results. Additional details on the super unit cells are provided in Table 2 and Table 3. For the rectangular PGUOM, as openness increases, both the unit cell width ($W_1 and W_3$) and slit width ($w_1 and w_3$) decrease. At an openness level of 0.73, the difference between $w_1$ and $w_3$ fall below 0.1 mm, which exceeds 3D printing resolution and present fabrication challenges. Achieving higher openness levels is possible with finer 3D printing resolution, allowing access to more precise design parameters for the PGUOM. Fig. 5(a) and 5(b) show the transmittance spectra and pressure field distribution of the rectangular PGUOM at different openness levels. In Fig. 5(a), our optimization process consistently achieves transmittance below 0.1 at 2000 Hz for openness levels from 0.4 to 0.73. Each openness level exhibits distinct broadband silencing characteristics, indicating no direct trade-off between openness and the effective working band. In Fig. 5(b), the transmission wave pressure substantially diminishes after passing through the rectangular PGUOM, reinforcing the sound silencing effect. Comparison with theoretical results in Fig. 2(k,l) shows the effectiveness of waveguide integration for optimization. Fig. 5(c) and 5(d) show the transmittance spectrum and pressure field distribution of the cylindrical PGUOM at different openness levels. The transmittance spectra reveal an effective silencing effect, and the pressure field distribution further confirms silencing at target frequency. 

\begin{figure}[h]
\centering
\includegraphics[width=\linewidth]{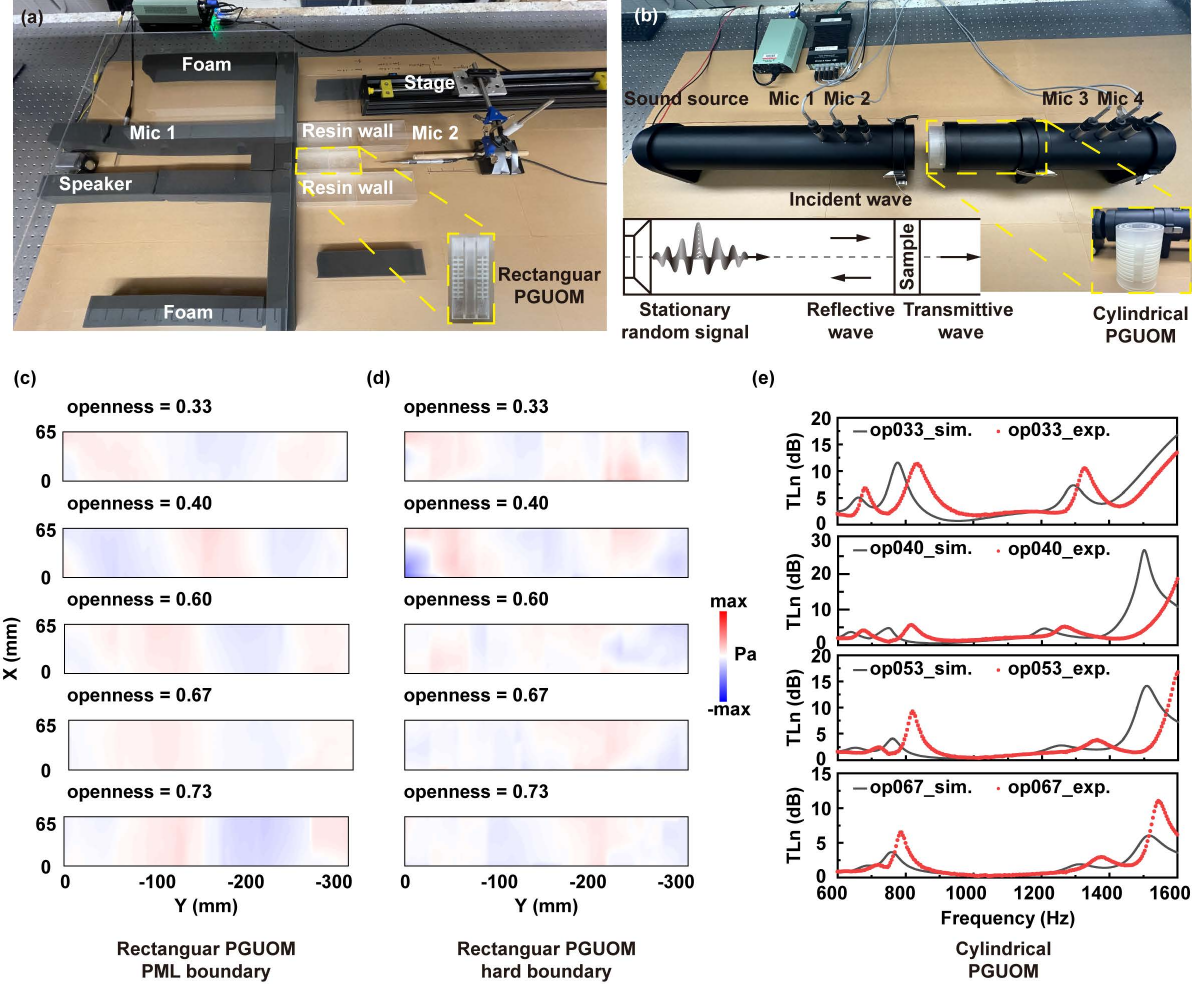}
\caption{(a-b) Illustration of the experimental for measuring the rectangular PGUOMs and cylindrical PGUOMs. (c-d) Measured pressure field distribution of transmitted wave through rectangular PGUOM under different openness levels, with PML boundary conditions and hard boundary conditions. (e) Simulated (black line) and measured (red dots) transmission loss of the cylindrical PGUOM across a frequency range from 600 Hz to 1600 Hz at different openness levels.}
\label{fig:6}
\end{figure}

\begin{table} [b]
\caption{Dimensions, transmission wave phases and amplitudes of the unit cells of rectangular PGUOM with different openness}
\centering
\begin{tabular}{ccccccccccc}
    \hline
$W_1/W_3/mm$ & $W_2/mm$ & $w_1/mm$ & $w_3/mm$ & $\varphi_1/\pi$ & $\varphi_2/\pi$ & $\varphi_3/\pi$ & $A_1$ & $A_2$ & $A_3$ & openness \\
    \hline
25 & 25 & 3.25 & 2.00 & 0.33 & & -0.33 & 0.69 & & 0.49 & 0.33 \\
22.5 & 30 & 2.61 & 1.78 & 0.30 & & -0.30 & 0.75 & & 0.56 & 0.40 \\
20 & 35 & 2.10 & 1.54 & 0.27 & & -0.26 & 0.79 & & 0.64 & 0.47 \\
17.5 & 40 & 1.63 & 1.30 & 0.23 & 1 & -0.22 & 0.85 & 1 & 0.73 & 0.53 \\
15 & 45 & 1.23 & 1.02 & 0.19 & & -0.21 & 0.89 & & 0.75 & 0.60 \\
12.5 & 50 & 0.89 & 0.78 & 0.18 & & -0.16 & 0.91 & & 0.84 & 0.67 \\
10 & 55 & 0.58 & 0.53 & 0.15 & & -0.12 & 0.93 & & 0.90 & 0.73 \\
    \hline
\end{tabular}
\label{tab:2}
\end{table}

\begin{table} [b]
\caption{Dimensions, transmission wave phases and amplitudes of the unit cells of cylindrical PGUOM with different openness}
\centering
\begin{tabular}{ccccccccccc}
    \hline
$R_1/mm$ & $R_2/mm$ & $\theta_1/\pi$ &$\theta_2/\pi$ & $\varphi_1/\pi $& $\varphi_2/\pi$ & $\varphi_3/\pi$ & $A_1$ & $A_2$ & $A_3$ & openness \\
    \hline
28.72 & 40.62 & 0.35 & 0.26 & 0.33 & 1 & -0.33 & 0.56 & 1 & 0.76 & 0.33 \\
27.25 & 41.62 & 0.7 & 0.179 & 0.3 & 1 & -0.30 & 0.75 & 1 & 0.56 & 0.40 \\
25.61 & 42.65 & 0.79 & 0.19 & 0.265 & 1 & -0.265 & 0.79 & 1 & 0.64 & 0.47 \\
24.12 & 43.51 & 0.49 & 0.19 & 0.235 & 1 & -0.235 & 0.85 & 1 & 0.73 & 0.53 \\
22.25 & 44.50 & 0.55 & 0.33 & 0.20 & 1 & -0.20 & 0.89 & 1 & 0.75 & 0.60 \\
20.21 & 45.46 & 0.69 & 0.20 & 0.165 & 1 & -0.165 & 0.91 & 1 & 0.84 & 0.67 \\
    \hline
\end{tabular}
\label{tab:3}
\end{table}

\section{Result and discussion}
To experimentally validate the PGUOM designs, we fabricated rectangular and cylindrical designs with varying openness levels and characterized their acoustic silencing performance. Since it is challenging to measure the transmittance of a rectangular PGUOM or the pressure field distribution of a cylindrical PGUOM, we opt to measure the pressure field distribution of the rectangular design and the transmission loss of a cylindrical design separately. Fig. 6(a-b) show the setup used to measure the pressure field distribution and transmission loss of the designs. Further details are provided in the method section. Fig. 6(c-d) display the pressure field distributions of rectangular PGUOMs with different openness level at 2000 Hz under PML boundary (foams) and hard boundary (resin walls). After passing through the design, the pressure of the transmitted wave decreases significantly and remains low throughout the remaining area, aligning well with the simulation results. Differences between the simulation and experimental results could be attributed to factors such as fabrication resolution and the sound absorption or reflection properties of the foam or the resin walls used in the experimental setup. 
Fig. 6(e) compares the simulated (black lines) and measured (red dots) transmission loss of cylindrical PGUOMs at various openness levels. For the simulations, thermoviscous losses close to the structure were considered, which decreases the peak transmission loss. The experimental results shows multiple transmission loss peaks for different designs. The observed difference between simulated results and measured results is primarily due to the connecting component added among unit cells during fabrication. Additional minor deviations could stem from fabrication inaccuracies in the experimental samples and slight simulation discrepancies in the acoustic properties.

\section{Conclusion}
In conclusion, this work presents a pioneering approach in the form of a phased gradient ultra open metamaterial, aiming to tackle the challenges associated with broadband, openness-adjustable, and ventilated sound silencing. The design, comprising three unit cells forming a phase gradient, effectively transforms incident waves into spoof surface waves at target frequency across different configurations. The significant reduction in sound pressure observed in the far field underscores the broad silencing band achieved by the PGUOM. Additionally, incorporating a blank, air-filled unit cell at the center of the super unit cell enables effective ventilation while maintaining strong sound silencing capabilities. Furthermore, the study demonstrates the adjustability of the openness without compromising the silencing efficiency. Numerical and experimental validations corroborate the effectiveness of the PGUOM design, with analytical, simulation, and experimental results aligning closely. Overall, the versatility of PGUOM, coupled with its ability to adapt to different boundary conditions and offer tunable broad working bands, rendering it suitable for a wide range of applications.

However, it’s important to acknowledge a limitation of the design, namely its sensitivity to the fabrication resolution of 3D printing. High target frequencies or large openness values may lead to dimensions approaching the limit of 3D printing resolution, potentially affecting the performance of the PGUOM. Nevertheless, with ongoing advancements in 3D printing materials \cite{sim20223d, zhang2022review, tan20223d}, there is considerable potential to enhance the design's functionalities and overcome fabrication challenges.

In summary, the PGUOM design represents a promising approach for addressing the complex requirements of broadband ventilated sound silencing, offering a versatile and adaptable solution with potential for further development and application in diverse fields.

\section{Materials and Methods}

\subsection{Numerical simulations}
All simulations were performed with finite element solver COMSOL Multiphysics using the pressure acoustic module in the frequency domain and Thermoviscous acoustic module in the frequency domain. Ports are used to provide incident with certain amplitude and phase. Rigid boundary, periodic boundary and perfectly matched layer boundary are used to represent boundary conditions under different situations. All unit cells are considered as perfectly rigid medium. At the reflection and transmission regions, perfectly matched layer has been implemented to mitigate the subsequent reflections.\break

\subsection{Fabrication and Characterization}
The PAUOMs were fabricated using the stereolithography (SLA) 3D printing technique. A commercial 3D printer (Formlabs Form 3+ SLA Printer) with a resolution of 25 microns was used. Two experimental setups are used to verify the attenuation effect for the planar design and cylindrical design respectively, as shown in the supporting information. In the case of the planar sample measurement, acrylic sheets of dimensions $61 cm×91 cm×0.6 cm$ and dimensions $30 cm×61 cm×0.2 cm$ placed in parallel with a spacing of 3cm to create reflection and transmission domains. Domain boundaries were confined with an absorbing foam 6cm in thickness to mitigate the back reflection. A loudspeaker was placed at one end of the reflection domain to generate a plane wave. One microphone was placed near the speaker and another microphone was fixed on a stage at the end of the sample. By adjusting the stage’s position, the microphone could measure the field pressure at various locations. The spacing between adjacent measurement locations is $3.125 mm×3.125 mm$. As for the cylindrical design, an impedance tube setup (100 mm B\&K 4206-T large impedance tube) was used to measure the transmission loss. The sample was placed in the sample holder, and four microphones are placed in the front and rear waveguides to assess the sample’s acoustic response. 

\section*{Acknowledgments}
The authors would like to thank Boston University Photonics Center and the Rajen Kilachand Fund for Integrated Life Science and Engineering for funding and technical support.

\subsection*{Author Contributions} 
X. Zhang, and Z. Yang conceived the study. Z. Yang, and X. Zhang conducted the numerical modeling and theoretical analysis, Z. Yang, X. Xie, and X. Zhang constructed the metamaterials. Z. Yang, A. Chen, and X. Zhang designed and conducted the experiments. All authors participated in discussing the results. X. Zhang, S. W. Anderson, and Z. Yang wrote the manuscript.

\subsection*{Funding}
This research was in part supported by the Boston University Photonics Center and the Rajen Kilachand
Fund for Integrated Life Science and Engineering.

\subsection*{Conflicts of Interest}
Provisional patent is filed based on the findings of this paper, detailed information is provided below. Provisional Patent Name: PHASED ARRAY ULTRA OPEN METAMATERIALS FOR BROADBAND ACOUSTIC SILENCING. Inventors: Xin Zhang, Zhiwei Yang, Ao Chen, Stephan Anderson; U.S. Provisional Application No.: 63/549,846; BU Ref.: BU-2024-010; BL Ref. No.: BOS-0072PR (56136.03184); File date: 02/05/2024.

\subsection*{Data Availability}
The authors declare that all data supporting the results of the study are available within the published article and its Supplementary Information section. All raw data generated during the current study are available from the corresponding author upon request.

\section*{Supplementary Materials}
Sections S1 to S6.
Figures. S1 to S8.

%\printbibliography
\bibliographystyle{plain}   % Choose a style
\bibliography{PGUOM}        % Add your .bib file
\end{document}